\documentclass[reprint,notitlepage,onecolumn,superscriptaddress,nofootinbib,amsmath,amssymb,aps,prx]{revtex4-1}

\usepackage[a4paper, margin=2.50cm]{geometry}

\usepackage{bbm}
\usepackage{graphicx}
\usepackage{fancyhdr, color}
\definecolor{myurlcolor}{rgb}{0,0,0.7}
\definecolor{myrefcolor}{rgb}{0.8,0,0}
\usepackage[unicode=true,pdfusetitle, bookmarks=false,bookmarksnumbered=false,
bookmarksopen=false, breaklinks=false,pdfborder={0 0 0},backref=false,
colorlinks=true, linkcolor=myrefcolor,citecolor=myurlcolor,urlcolor=myurlcolor]{hyperref}

\renewcommand{\H}{\hat{H}}

\newcommand{\ee}{\ensuremath{\mathrm{e}}}

\newcommand{\bra}[1]{\ensuremath{\left\langle #1 \right\rvert}}
\newcommand{\ket}[1]{\ensuremath{\left\lvert #1 \right\rangle}}

\newcommand{\Tr}{\mathrm{Tr}}

\renewcommand{\a}{\ensuremath{\hat{a}}}
\newcommand{\adag}{\ensuremath{\hat{a}^\dagger}}

\newcommand{\sg}{\ensuremath{\hat{\sigma}}}
\newcommand{\sgd}{\ensuremath{\hat{\sigma}^\dagger}}

\renewcommand{\H}{\ensuremath{\hat{H}}}

\newcommand{\LL}{\ensuremath{\mathcal{L}}}

\newcommand{\DD}{\ensuremath{\mathcal{D}}}

\newcommand{\eqr}[1]{Eq.~\eqref{#1}}

\begin{document}

\fancyhead[C]{\sc \color[rgb]{0.4,0.2,0.9}{Physical implementations of quantum absorption refrigerators}}
\fancyhead[R]{}
\fancyhead[L]{}

\title{Physical implementations of quantum absorption refrigerators}

\author{Mark T. Mitchison}
\email{markTmitchison@gmail.com} 
\affiliation{Institut f\"ur Theoretische Physik, Albert-Einstein Allee 11, Universit\"at Ulm, D-89069 Ulm, Germany}

\author{Patrick P. Potts}
\email{patrick.hofer@teorfys.lu.se} 
\thanks{formerly known as Patrick P. Hofer}
\affiliation{Department of Applied Physics, University of Geneva, Chemin de Pinchat 22, 1211 Geneva, Switzerland.}
\affiliation{Physics Department and NanoLund, Lund University, Box 118, 22100 Lund, Sweden.}

\date{\today}

\begin{abstract}
Absorption refrigerators are autonomous thermal machines that harness the spontaneous flow of heat from a hot bath into the environment in order to perform cooling. Here we discuss quantum realizations of absorption refrigerators in two different settings: namely, cavity and circuit quantum electrodynamics. We first provide a unified description of these machines in terms of the concept of virtual temperature. Next, we describe the two different physical setups in detail and compare their properties and performance. We conclude with an outlook on future work and open questions in this field of research.
\end{abstract}

\maketitle


The investigation of quantum thermal machines is a subfield of quantum thermodynamics which is of particular practical relevance. A thermal machine is a device which utilizes a temperature bias to achieve some useful task (for recent reviews see e.g.~\cite{kosloff:2014,gelbwaser:2015,goold:2016,vinjanampathy:2016,benenti:2017,kosloff:2013, alicki:intro} as well as Ref.~\cite{wholebook}, Part I). The nature of this task can vary, with the most prominent example being the production of work in heat engines. Further examples include the creation of entanglement~\cite{brask:2015njp}, the estimation of temperature \cite{hofer:2017} and the measurement of time~\cite{erker:2017}. 

This chapter discusses physical implementations of quantum absorption refrigerators. There are two compelling reasons which motivate the pursuit of implementing exactly this thermal machine. First, the absorption refrigerator is an autonomous thermal machine. Autonomous machines only require static biases of temperature and/or chemical potential and do not require any time-dependent control. Therefore, they dispense of any controversy over the definition of work in the quantum regime (see also Ref.~\cite{wholebook}, Part II) as well as the energetic cost of maintaining control over time-dependent fields \cite{clivaz:2017,Woods2016,erker:2017}. Furthermore, autonomous machines connect to naturally occurring processes (e.g., photosynthesis~\cite{Dorfman2013,Creatore2013,Killoran2015}), where no external control is needed to harness energy. They are thus promising candidates for harvesting waste heat from other processes or naturally abundant (thermal) energy sources. The second reason why the quantum absorption refrigerator is of great practical relevance is the fact that it is a refrigerator. As discussed below, the quantum absorption refrigerator can stabilize single degrees of freedom in low-entropy states. The ability to initiate a quantum system in a low-entropy state is crucial for most experiments in the quantum regime. In addition to performing state preparation, implementing a quantum absorption refrigerator allows for investigating the fundamental physics of this important process, illuminating its capabilities and limitations. 

Since the initial theoretical conception of quantum absorption refrigerators~\cite{palao:2001,linden:2010prl,levy:2012}, several experimental proposals have been put forward~\cite{Chen2012,mari:2012,venturelli:2013,mitchison:2016,hofer:2016}, with one experiment having realized a quantum absorption refrigerator in a trapped-ion system~\cite{maslennikov:2017}. Here, we focus on the setups described in Refs.~\cite{hofer:2016,mitchison:2016}, in the context of cavity and circuit quantum electrodynamics (QED). These involve different physical degrees of freedom and operate in distinct regimes of energy, temperature, and cooling performance. Nevertheless, as we shall explain, both can be understood in terms of the concept of \textit{virtual temperature}~\cite{brunner:2012,skrzypczyk:2015}, which provides a unifying framework for understanding autonomous thermal machines. 

The rest of this chapter is structured as follows: In Sec.~\ref{sec:QAR}, the theory of the quantum absorption refrigerator and the relevant concepts and mechanisms are reviewed. Physical implementations in the setting of cavity and circuit QED are then discussed in Secs.~\ref{sec:atom_cavity} and \ref{sec:cp}. The chapter concludes with a comparison of the two implementations and an outlook in Sec.~\ref{sec:comparison}.

\section{The quantum absorption refrigerator}
\label{sec:QAR}
Unlike the refrigerator found in a typical kitchen, an absorption refrigerator does not rely on energy in the form of mechanical work. Instead, an absorption refrigerator harnesses the spontaneous flow of heat from a hot bath into the environment in order to perform cooling. Let us first analyze the thermodynamics of a generic absorption refrigerator before introducing its quantum realization. 

\begin{figure}
\includegraphics[scale=0.3]{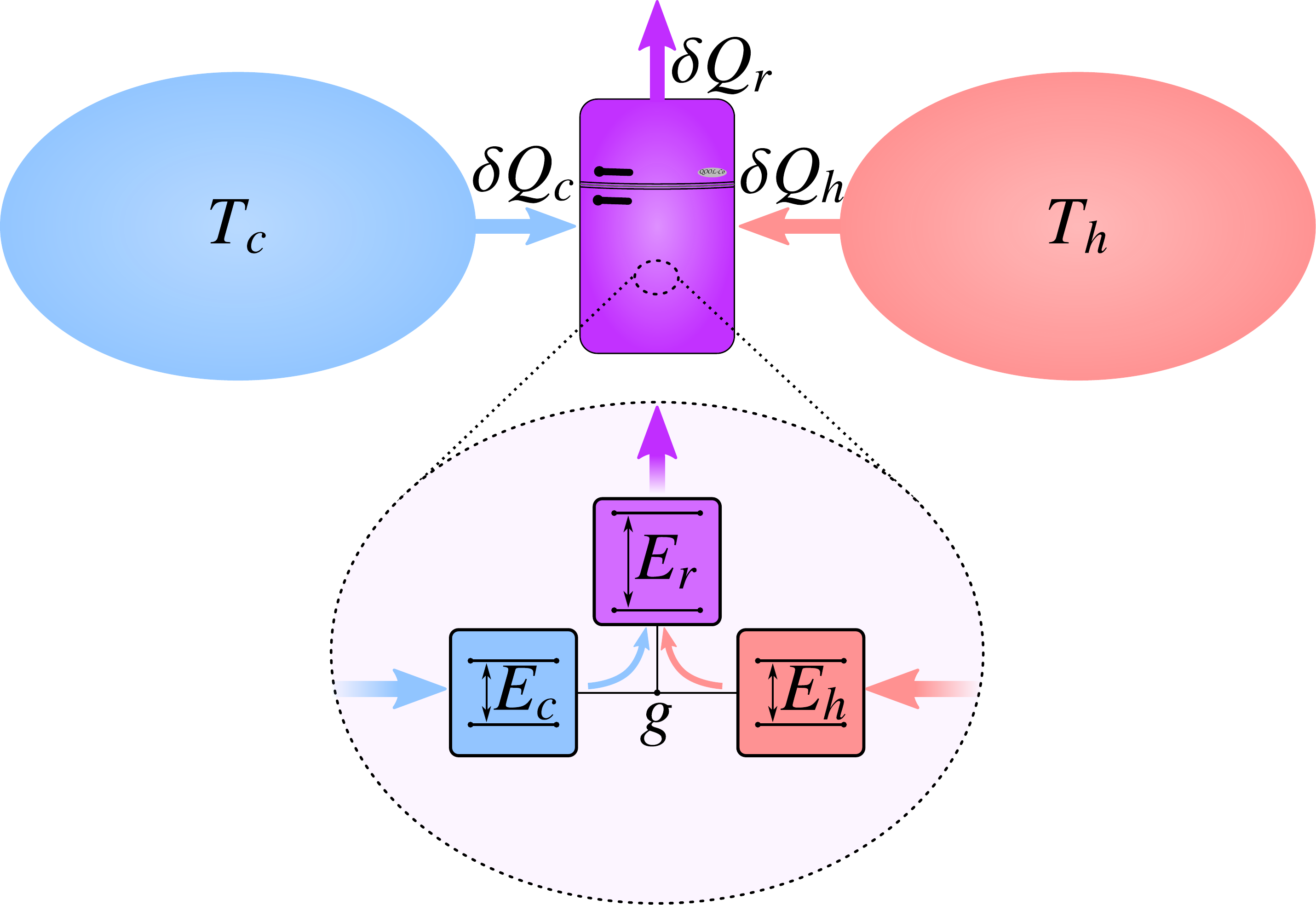}
\caption{An absorption refrigerator removes heat simultaneously from a hot ($T_h$) and a cold ($T_c$) body and dumps it into the environment at ambient temperature $T_r$. We consider quantum absorption refrigerators comprising three subsystems with transitions of energy $E_c$, $E_h$ and $E_r = E_c + E_h$, coupled by a three-body interaction of strength $g$.\label{absorption_fridge}}
\end{figure}

An absorption refrigerator interacts with three thermal reservoirs: a cold one at temperature $T_c$ representing the body to be cooled, a reservoir at room temperature $T_r$ representing the environment, and a hot reservoir at temperature $T_h$ which provides a source of free energy (see Fig.~\ref{absorption_fridge}), with $T_c\leq T_r < T_h$. Suppose that the refrigerator absorbs quantities of heat $\delta Q_h$ from the hot reservoir and $\delta Q_c$ from the cold reservoir, while dumping $\delta Q_r$ heat into the environment. Under steady-state conditions, and assuming the ideal case where the internal mechanism of the refrigerator is frictionless and without heat leaks \cite{correa:2015pre}, the first law of thermodynamics reads as
\begin{equation}
\label{FirstLaw}
\delta Q_c + \delta Q_h = \delta Q_r.
\end{equation}
If we also assume idealized heat reservoirs then the total entropy change of this process is 
\begin{equation}
\label{SecondLaw}
\delta S = \frac{\delta Q_r}{T_r} - \frac{\delta Q_h}{T_h} - \frac{\delta Q_c}{T_c} \geq  0.
\end{equation}
According to the second law of thermodynamics, such a process is physically permissible so long as the entropy change is non-negative.

In order to design a \textit{quantum} absorption refrigerator, we need to model a dynamical process that transfers heat between three reservoirs in the manner described above. A simple example of such a model comprises three components or subsystems~\cite{linden:2010prl,levy:2012}, referred to as $c$, $r$ and $h$, each of which is connected to one of the three reservoirs. We use the subscripts $c$, $r$ or $h$ to denote quantities associated to the subsystem coupled to the corresponding bath.

The refrigerator Hamiltonian $\nolinebreak{\H = \sum_j \H_j + \H_{\rm int}}$ can be resolved into the local energy $\H_j$ of each component, for $j = c,r,h$, together with the interaction between them, $\H_{\rm int}$. We assume that the energy spectrum of each subsystem consists of an equally spaced (possibly infinite) ladder of states separated in energy by an amount $E_j$, i.e.
\begin{equation}
\label{localHamiltonianLadder}
\H_j = E_j \sum_{n} n\ket{n}_j\bra{n},
\end{equation}
where the integer $n\geq 0$ labels the energy eigenstates. Such a Hamiltonian describes, for example, a two-level atom, a spin in a static magnetic field, or a harmonic oscillator. The interaction between the subsystems is governed by the three-body Hamiltonian
\begin{equation}
\label{threeBodyInteraction}
\H_{\rm int} = g \left ( \hat{L}_c \hat{L}^\dagger_r \hat{L}_h + \hat{L}_c^\dagger \hat{L}_r \hat{L}_h^\dagger \right ).
\end{equation}
Here, $g$ is a coupling constant that is assumed to be small in comparison to the local energies $E_j$, while $\hat{L}_j$ is a lowering operator with respect to the corresponding local Hamiltonian, i.e.
\begin{equation}
\label{loweringOperator}
[\H_j,\hat{L}_j] = -E_j \hat{L}_j.
\end{equation}
In other words, the nonzero matrix elements of $\hat{L}_j$ connect states $\ket{n}_j$ and $\ket{n+1}_j$ separated in energy by $E_j$. Therefore, the interaction~\eqref{loweringOperator} induces transitions between the states 
\begin{equation}
\label{resonantTransition}
\ket{l+1}_c\ket{m}_r\ket{n+1}_h\longleftrightarrow \ket{l}_c\ket{m+1}_r\ket{n}_h.
\end{equation}
This provides a channel for energy to flow from subsystem $h$ to $r$, but only by simultaneously extracting some energy from $c$. At the same time, each component of the refrigerator continuously absorbs or dissipates heat due to its contact with the corresponding bath. The net effect is a cooperative flow of heat from the cold and hot reservoirs into the room-temperature environment.

In order for $\H_{\rm int}$ to have an appreciable effect for small $g$, the transition~\eqref{resonantTransition} should be resonant, which is enforced by the condition
\begin{equation}
\label{resonanceCondition}
E_c + E_h = E_r.
\end{equation}
Since $E_j$ is the smallest quantity of heat which can be absorbed by each subsystem, this represents a direct analogue of the first law~\eqref{FirstLaw} at the level of individual energy quanta. Note that here, in identifying $\delta Q_j = E_j$, we are again implicitly assuming steady-state operation and the absence of heat leaks in the transfer of energy between the baths mediated by the interaction~\eqref{threeBodyInteraction}, which is valid so long as the coupling $g$ is sufficiently weak~\cite{correa:2013,hofer:2017njp,Gonzalez2017,seah:2018}. If we similarly set $\delta Q_j = E_j$ in the second law~\eqref{SecondLaw}, we obtain the condition 
\begin{equation}
\label{virtualTemperature}
T_c \geq\frac{E_r - E_h}{E_r/T_r- E_h/T_h}\equiv T_v.
\end{equation}
Thus, the ability of the refrigerator to cool is determined by a new temperature scale, $T_v$, the \textit{virtual temperature}~\cite{brunner:2012}.

\begin{figure}
\includegraphics[scale=0.3]{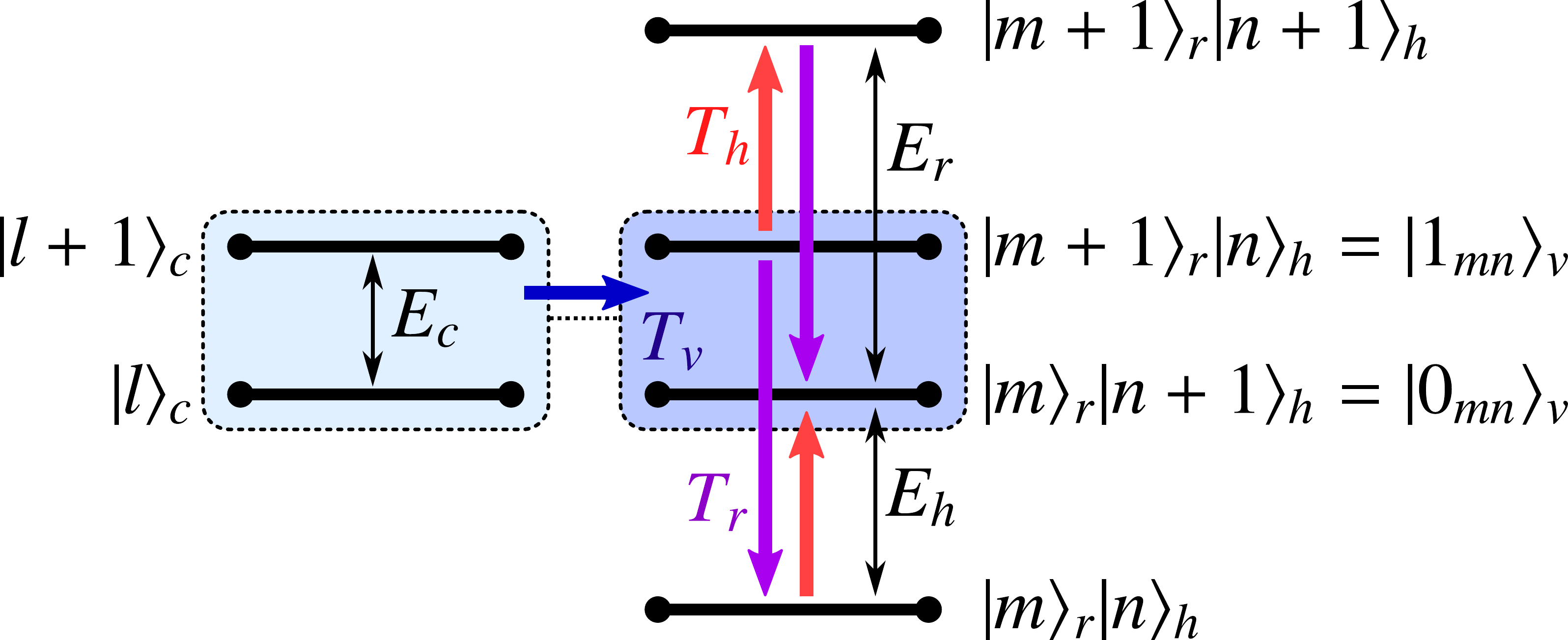}
\caption{A virtual qubit is a pair of states in the composite Hilbert space of a quantum absorption refrigerator, which is resonant with the target object to be cooled. Transitions driven by physical baths at temperatures $T_r$ and $T_h$ drive population into the lower-energy virtual-qubit state. The resulting population imbalance is described by a cold virtual temperature $T_v$, thus inducing heat to flow out of the target.
\label{virtual_qubit}}
\end{figure}

In order to understand what the virtual temperature represents microscopically, consider the pairs of states $\ket{0_{mn}}_v = \ket{m}_r\ket{n+1}_h$ and $\ket{1_{mn}}_v=\ket{m+1}_r\ket{n}_h$, each of which constitutes a \textit{virtual qubit}. The difference in energy between the virtual qubit's states is $E_r-E_h = E_c$ by virtue of the constraint~\eqref{resonanceCondition}. Suppose that initially the interaction is switched off and each subsystem equilibrates with its corresponding bath. The equilibrium populations $p^{(j)}_n$ of the energy eigenstates $\ket{n}_j$ of each subsystem are given by the Boltzmann distribution, so that
\begin{equation}
\label{populationRatio}
\frac{p_{n+1}^{(j)}}{p^{(j)}_n} = \ee^{-E_j/k_B T_j}.
\end{equation}
It follows immediately that the virtual qubit populations are Boltzmann-distributed at the virtual temperature, i.e.,
\begin{equation}
\label{virtualQubitRatio}
\frac{p^{(v)}_{1_{mn}}}{p_{0_{mn}}^{(v)}} = \frac{p_{m+1}^{(r)} p_{n}^{(h)}}{p_{m}^{(r)} p_{n+1}^{(h)}} = \ee^{-E_c/k_B T_v}.
\end{equation}
Now, switching on the interaction $\H_{\rm int}$ causes resonant transitions according to Eq.~\eqref{resonantTransition} or, equivalently,
\begin{equation}
\label{virtualQubitTransitionScheme}
\ket{l+1}_c\ket{0_{mn}}_v\longleftrightarrow \ket{l}_c\ket{1_{mn}}_v,
\end{equation}
leading to thermalization of $c$ with the virtual qubits. Meanwhile, exchange of heat with the hot and room-temperature reservoirs acts to maintain the virtual qubits at temperature $T_v$ (see Fig.~\ref{virtual_qubit}). A net heat flux out of $c$ is therefore established so long as the virtual qubits are effectively colder, i.e., if $T_c \geq T_v \geq 0$, which is precisely the result obtained in Eq.~\eqref{virtualTemperature} from purely thermodynamic arguments. This condition is also sometimes equivalently expressed in terms of the cooling window for the energy splitting $E_c$,
\begin{equation}
\label{coolingWindow}
0 \leq E_c \leq  \frac{(T_h-T_r)T_c}{(T_r-T_c)T_h}E_h.
\end{equation}

In order to analyze the performance of the refrigerator, we need to choose appropriate figures of merit according to the specific task at hand.  One possibility is that, by analogy with the classical case, the purpose of the refrigerator is to extract heat from the \textit{macroscopic bath} at temperature $T_c$. In this case, performance can be measured in terms of the steady-state heat currents $J_j = \partial Q_j/\partial t$, which we define as the rate at which heat flows \textit{into} the system from baths $j = c,h$, and the rate at which heat flows \textit{out of} the system to the room-temperature bath for $j = r$. The cooling power is defined by $J_c$, while the coefficient of performance (COP) $\varepsilon = J_c/J_h$ gives a measure of efficiency. We assume, as above, that each quantity of heat $E_r$ dumped into the environment corresponds to precisely one quantum of energy $E_c$ and $E_h$ absorbed from each of the cold and hot baths, respectively. Therefore, the heat currents must satisfy $J_j = \chi E_j$, with $\chi$ a common proportionality factor, and we obtain the very simple expression
\begin{equation}
\label{COPperfect}
\varepsilon = \frac{E_c}{E_h}.
\end{equation}
This universal result is valid under the assumption of ideal heat transfer between the baths \cite{scovil:1959}.

Combining the first and second laws of thermodynamics, Eqs.~\eqref{FirstLaw} and \eqref{SecondLaw}, one obtains the Carnot bound~\cite{palao:2001,skrzypczyk:2011}
\begin{equation}
\label{CarnotCOP}
\varepsilon \leq \varepsilon_C = \frac{1-T_r/T_h}{T_r/T_c-1}.
\end{equation}
Note that this bound is consistent with Eqs.~\eqref{coolingWindow} and \eqref{COPperfect}. The Carnot point $\varepsilon=\varepsilon_C$ corresponds to reversible operation, with $\delta S = 0$. The cooling power is clearly zero at the Carnot point because $T_v = T_c$ and thus the effective temperature difference driving the heat current vanishes. Note that, unlike the efficiency of a heat engine, the COP can be greater than unity. Indeed, in the limit $T_r\to T_c$, we have that $\varepsilon_C \to \infty$ and the COP can be arbitrarily large. This makes sense, since heat can be transferred between two baths at equal temperatures without entropic cost, in principle.

A different approach is to consider the refrigerator as a machine designed to cool down the \textit{quantum system} $c$ itself. In this case, the achievable temperature is a more relevant figure of merit. More generally, since the final state of $c$ will generally be out of equilibrium, one can consider the achievable energy or entropy instead. This situation is frequently encountered in practice, since the objective of refrigeration is often to \textit{prepare} the quantum system in a low-entropy state as a precursor to some experiment. It is clearly additionally advantageous to reach the minimum temperature as quickly as possible, which opens up interesting possibilities for engineering the transient behavior of the refrigerator to gain some enhancement. In particular, since the dynamics induced by the interaction~\eqref{threeBodyInteraction} may be oscillatory, the minimum temperature may be reached in a finite time~\cite{brask:2015,mitchison:2015}. In order for this to be useful, there must exist some mechanism to switch off the interaction at the appropriate moment. The superconducting-circuit refrigerator discussed in Sec.~\ref{sec:cp} represents an example where this is the case.

We finally mention that, by a judicious choice of the physical energies and temperatures, the virtual temperature can be made to take any desired value. For example, if $T_v > T_c$, then the system operates as a heat pump that transfers energy from $r$ to $h$. Indeed, even negative values for $T_v$ are possible. In this case, the dynamics induces population inversion in the states of subsystem $c$, raising its energy in an analogous manner to a heat engine lifting a weight. If the autonomous machine acts on a more complex system that possesses multiple transitions at different energies, then the machine itself must also be more complex, having at least one virtual qubit (and associated virtual temperature) resonantly coupled to each of these transitions. The most efficient machines (the ones reaching Carnot efficiency) make use of a single virtual temperature only \cite{brunner:2012}. Therefore, the virtual temperature provides a unified description of autonomous quantum thermal machines, from refrigerators to engines.


\section{Optomechanical coupling in cavity QED}
\label{sec:atom_cavity}
One important application of refrigeration in the quantum regime is the cooling of atomic motion. Indeed, the development of laser-cooling techniques laid the foundation for much of the research on controlled quantum dynamics that is carried out today. This includes the field of cavity QED, which is a mature experimental platform for investigating the interaction of light and matter at the atomic level (see, for example, Refs.~\cite{Miller2005,Walther2006,Haroche2013} and references therein). In this section, we will discuss a physical realization of a quantum absorption refrigerator designed to cool the motion of a trapped atom inside two optical cavities~\cite{mitchison:2016}. The scheme is very closely related to conventional laser sideband cooling~\cite{Leibfried2003}, which allows us to draw a connection between modern developments in quantum thermodynamics and standard refrigeration methods in atomic physics.

\subsection{Origin of the optomechanical interaction}

\begin{figure}
\includegraphics[scale=0.5]{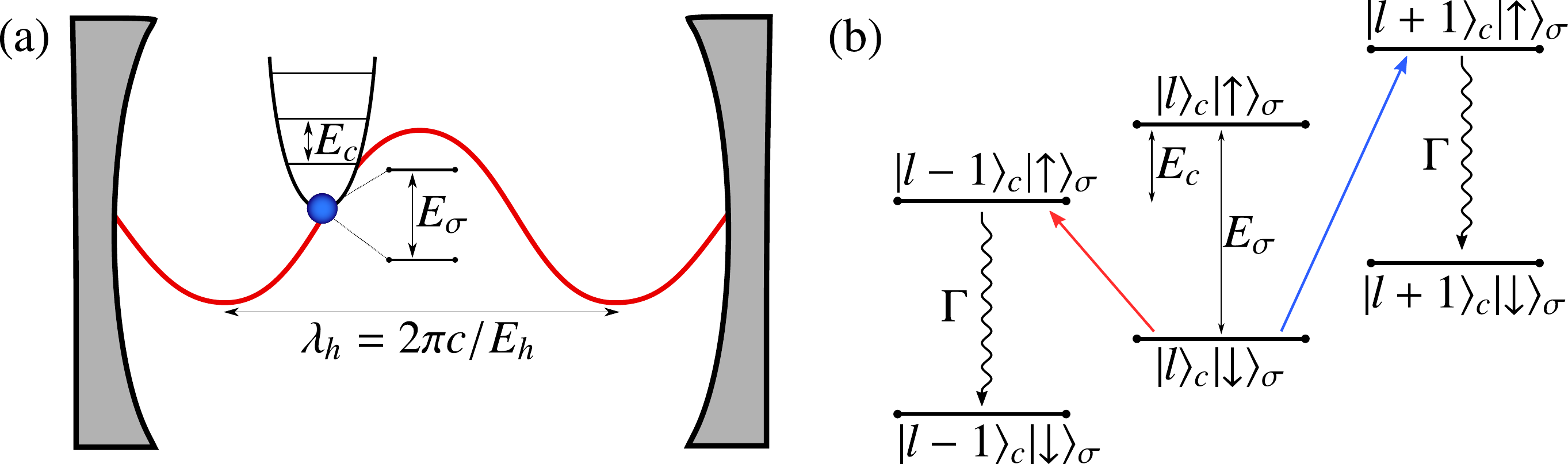}
\caption{(a) Trapping a two-level atom near the node of a cavity mode gives rise to an optomechanical coupling. (b) The atomic motion splits the electronic transition into sidebands separated by frequency $E_c$. Red and blue arrows show the corresponding (first) sideband transitions. Driving transitions on the red sideband leads to cooling by converting phonons into spontaneously emitted photons (wavy line), while blue sideband transitions lead to heating. \label{single_cavity}}
\end{figure}

To illustrate the physical principles underlying the atom-cavity refrigerator, we first consider a two-level atom or ion of mass $m$ confined by a harmonic trap that is placed inside an optical cavity aligned parallel to the $x$ axis, as depicted in Fig.~\ref{single_cavity} (a). Let us focus on a particular vibrational mode with angular frequency $E_c$, corresponding to motion parallel to the cavity axis, and an optical mode with angular frequency $E_h$ and wavelength $\lambda_h = 2\pi c/E_h$, with $c$ the speed of light. We set $\hbar =1$ throughout. The internal electronic levels $\ket{\uparrow}_\sigma$ and $\ket{\downarrow}_\sigma$ are separated by an energy $E_\sigma$.  The free Hamiltonian is thus
\begin{equation}
\label{bareHamiltonianSingle}
\H_0 =  E_c \adag_c\a_c  + E_\sigma \sgd\sg + E_h \adag_h\a_h.
\end{equation}
Here, $\a_c$ and $\a_h$ are bosonic ladder operators that annihilate vibrational quanta (phonons) and cavity photons, respectively, while $\sg = \ket{\downarrow}_\sigma\bra{\uparrow}$ is a lowering operator for the electronic levels. For typical vibrational and optical frequencies, we have that $E_c\ll E_\sigma, E_h$. Taking the electronic and vibrational degrees of freedom together, the spectrum of transition frequencies of the atomic system consists of a central (carrier) frequency $E_\sigma$ split into closely spaced sidebands separated in frequency by $E_c$. 

The light-matter interaction Hamiltonian in the dipole approximation is given by $\H_{\rm int} = -\hat{\bf d}\cdot \hat{\bf E}(\hat{\bf r}) $, with $\hat{\bf d}$ the dipole moment operator and $\hat{\bf E}(\hat{\bf r})$ the electric field operator of the cavity mode evaluated at the atom's center of mass $\hat{\bf r} = (\hat{x},\hat{y},\hat{z})$. Neglecting the slow spatial variation of the field in the $y$ and $z$ directions, we have~\cite{Blockley1992, Buzek1997}
\begin{equation}
\label{dipoleInteractionSingle}
\H_{\rm int} = g_0 \sin\left (\frac{2\pi\hat{x}}{\lambda_h}\right)\left (\sg\adag_h + \sgd\a_h\right ),
\end{equation}
where the cavity coupling constant $g_0$ is determined by the product of the transition dipole moment and the electric-field strength per photon. Here we have made a rotating-wave approximation by neglecting counter-rotating terms such as $\sgd\adag_h$, which is valid so long as $g_0\ll E_\sigma + E_h$. 

The center-of-mass coordinate can be written as $\hat{x} = x_0 + x_{\rm zp}(\a_c+\adag_c)$, with $x_0$ the minimum of the trap potential and $x_{\rm zp} = 1/\sqrt{2mE_c}$ the width of the vibrational ground-state wavefunction (zero-point motion). The optomechanical interaction is then characterized by the dimensionless phase $\phi = 2\pi x_0/\lambda_h$ and Lamb-Dicke parameter $\eta = 2\pi x_{\rm zp}/\lambda_h = E_h/\sqrt{mc^2E_c}$. In particular, we may write $\H_{\rm int} = \cos(\phi) \H_{\rm node} + \sin(\phi)\H_{\rm anti}$, where 
\begin{align}
\label{Hnode}
\H_{\rm node} & = g_0\sin \left[\eta \left (\a_c+\adag_c\right )\right ]\left (\sg\adag_h + \sgd\a_h\right ),\\
\label{Hanti}
\H_{\rm anti} & = g_0\cos \left[\eta \left (\a_c+\adag_c\right )\right ]\left (\sg\adag_h + \sgd\a_h\right ).
\end{align}
These two Hamiltonians describe the light-matter interaction when the minimum of the trap potential coincides either with a node of the cavity field, where $\phi = n\pi$ for integer $n$, or with an anti-node, where $\phi = (n+\tfrac{1}{2})\pi$. 

At the node, we see that the emission and absorption of cavity photons must be associated with a change in the motional state. This can be understood from symmetry: The electric field switches sign under a parity transformation $(\hat{x}-x_0) \to -(\hat{x}-x_0)$, and therefore the matrix elements of $\H_{\rm int}$ vanish between states of the same parity. Furthermore, the motional energy eigenstates $\ket{l}_c$, with $l$ the integer phonon number, have definite parity $(-1)^{l}$. This gives rise to a selection rule $\Delta l = 2m-1$, where the state must change by an odd number of phonons. Conversely, at the anti-node the electric field is parity-invariant and thus couples states with the same parity, leading to the selection rule $\Delta l = 2m$.

Typically, $\lambda_h\lesssim 1\,\mu \rm m$ while $x_{\rm zp}\lesssim 10 \,\rm nm$, so that $\eta$ is a small parameter. In the Lamb-Dicke regime where $\eta \sqrt{n_c} \ll 1$, with $n_c = \langle \adag_c\a_c\rangle$ the mean phonon number, we may expand the Hamiltonian up to first order in $\eta$ (Lamb-Dicke approximation). First, consider the case where the potential minimum is placed at a node and the cavity frequency is tuned to the first red sideband, $E_h = E_\sigma - E_c$. Neglecting terms that do not commute with $\H_0$, the interaction Hamiltonian can be approximated as $\H_{\rm node}\approx \H_{\rm red}$, where
\begin{equation}
\label{Hred}
\H_{\rm red} = g_0\eta \left (\a_c\sgd\a_h + \adag_c\sg\adag_h \right ).
\end{equation}
On the other hand, tuning the cavity to the first blue sideband, $E_h = E_\sigma + E_c$, we similarly obtain $\H_{\rm node} \approx \H_{\rm blue}$, with
\begin{equation}
\label{Hblue}
\H_{\rm blue} = g_0\eta \left (\adag_c\sgd\a_h + \a_c\sg\adag_h \right ).
\end{equation}
Meanwhile, at the anti-node, to lowest order in $\eta$ we find $\H_{\rm anti}\approx \H_{\rm JC}$, with the Jaynes-Cummings Hamiltonian~\cite{Jaynes1963,Shore1993}
\begin{equation}
\label{Hcar}
\H_{\rm JC} = g_0\left(\sgd\a_h + \sg\adag_h \right ).
\end{equation}
Thus, the type of optomechanical interaction can be selected by shifting the trap potential and the cavity frequency. Note, however, that shifting the trap potential directly would create motional excitations unless done very slowly, i.e., adiabatically. A simple dimensional estimate yields an upper velocity bound $\partial x_0/\partial t  \ll x_{\rm zp} E_c/2\pi \sim 10~\mu{\rm m/ms}$. On the other hand, shifting the cavity itself could be done much faster without creating excitations, due to the higher optical frequencies. Nevertheless, the system described in Sec.~\ref{sec:cp} appears to be a more promising setting in which to implement such switching of the interaction.

The red sideband Hamiltonian $\H_{\rm red}$ is a three-body interaction of the required form~\eqref{threeBodyInteraction} to implement a quantum absorption refrigerator. The thermal baths could be realized by driving the cavity with thermal light at high temperature $T_h$, while the electronic degrees of freedom naturally couple to the external radiation field at room temperature $T_r$. The cooling mechanism can be understood by analogy with laser sideband cooling, but with the coherent laser field replaced by a hot cavity mode. In sideband cooling, photons on the first red sideband are resonantly absorbed, thus transferring energy from the motional to internal atomic degrees of freedom [cf.~Eq.~\eqref{Hred}]. The excess energy is then lost to the environment by spontaneous emission, which resets the electronic state so that it is ready to absorb another photon-phonon pair. This drives population down the ladder of motional states, as illustrated in Fig.~\ref{single_cavity} (b). 

Unfortunately, this picture in terms of distinct sideband transitions is spoiled by the large natural linewidth of a typical electronic transition associated with a fast spontaneous emission rate $\Gamma \gg E_c$. The resulting energy uncertainty means that the motional sidebands are not ``resolved'', i.e., they cannot be efficiently and selectively addressed by tuning the cavity frequency. Therefore, the rotating-wave approximation leading to Eq.~\eqref{Hred} breaks down. As a result, Eq.~\eqref{Hblue} also contributes significantly to the dynamics by driving blue sideband transitions, which increase the number of phonons and therefore lead to motional heating (see Fig.~\ref{single_cavity}). The momentum recoil associated with spontaneous emission leads to further heating and thus represents an additional drawback of this setup.

In the context of laser sideband cooling, these issues can be circumvented by introducing a second laser field and driving red-sideband transitions by a two-photon Raman process~\cite{Leibfried2003}, which entirely avoids populating the excited electronic state. A similar strategy can be adopted here by introducing a second optical cavity, as explained in the following section.

\subsection{Crossed-cavity absorption refrigerator}

\begin{figure}
\includegraphics[scale=0.5]{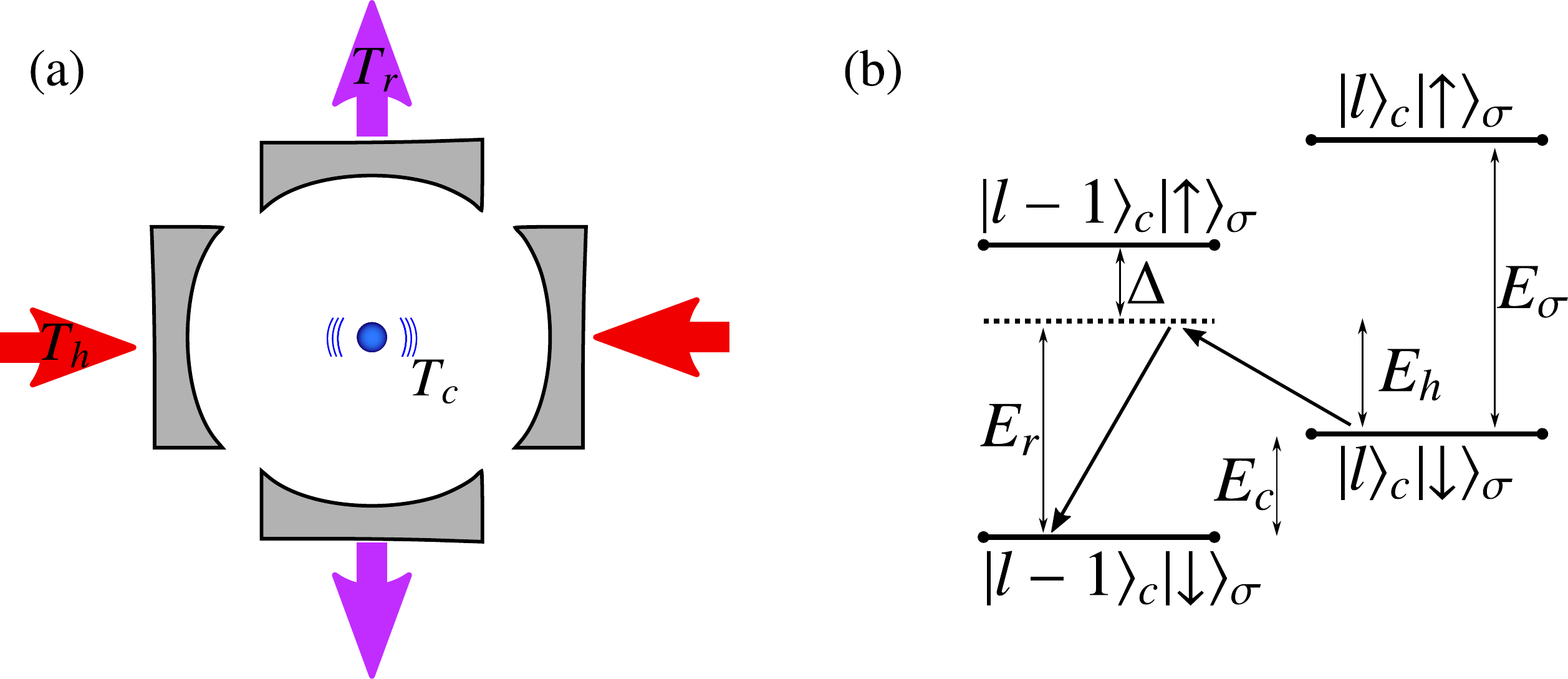}
\caption{(a) Illustration of the crossed-cavity absorption refrigerator. An atom trapped near the node of one cavity and the anti-node of the other is cooled by driving one of the cavities with thermal light, while the other cavity couples to the room-temperature radiation field. (b) Cooling results from coherent transitions between motional states driven by the exchange of photons between the two cavities. \label{double_cavity}}
\end{figure}

In order to suppress spontaneous emission, we introduce a second cavity lying perpendicular to the first one, as illustrated in Fig.~\ref{double_cavity}. The free Hamiltonian now reads as
\begin{equation}
\label{bareHamiltonianDouble}
\H_0 = \sum_{j=c,r,h} E_j \adag_j\a_j + E_\sigma \sgd\sg.
\end{equation}
Here, we introduced a bosonic lowering operator $\a_r$ for the second cavity mode with frequency $E_r = E_h + E_c$. The minimum of the confining potential is positioned to coincide with a node of the first cavity and an anti-node of the second cavity. According to the discussion in the previous section, the interaction Hamiltonian in the Lamb-Dicke approximation reads as 
\begin{equation}
\label{LDinteraction}
\H_{\rm int} = g_h \eta \big (\a_c + \adag_c \big )\big (\sg\adag_h + \sgd\a_h\big ) + g_r \big (\sg\adag_r + \sgd\a_r\big ),
\end{equation}
where $g_{h,r}$ are the corresponding cavity coupling constants.

Let $\Delta = E_\sigma - E_r$ denote the detuning from electronic resonance, and $\Gamma$ the spontaneous emission rate. So long as $\Gamma \ll |\Delta|$ and $g_h g_r\eta \ll \Delta^2$, the excited electronic state is never appreciably populated and can be adiabatically eliminated from the dynamics. As shown in Ref.~\cite{mitchison:2016}, this leads to an effective Hamiltonian of the form
\begin{equation}
\label{Heff}
\H = \sum_{j=c,r,h} E_j \adag_j\a_j + g\left (\a_c\adag_r\a_h + \adag_c \a_r\adag_h \right ),
\end{equation}
with $g = -g_h g_r\eta/\Delta$ the effective coupling constant. This Hamiltonian, which has an interaction of the form~\eqref{threeBodyInteraction}, describes the coherent exchange of photons between the cavities, assisted by a change in phonon number to make up the energy mismatch $E_c = E_r-E_h$ between the cavity modes. Therefore, when mode $h$ is at a higher temperature than mode $r$, this exchange of photons drives red sideband transitions in a manner analogous to Raman sideband cooling, as illustrated in Fig.~\ref{double_cavity} (b). 

In order to implement the thermal baths, we assume that cavity $h$ couples to a thermal light field at temperature $T_h$, while cavity $r$ couples to another field at temperature $T_r$. Meanwhile, some intrinsic motional heating of the trapped atom is unavoidable, e.g., due to fluctuations of the confining potential. The ensuing dynamics of the three harmonic modes can be described by a master equation in Lindblad form
\begin{equation}
\label{masterEquationCavity}
\partial_t\hat{\rho} = -i[\H, \hat{\rho}] + \sum_{j=c,h,r}\LL_j \hat{\rho},
\end{equation}
where the superoperators $\LL_j$ describe the coupling to the corresponding bath. We adopt a local description of dissipation, as appropriate for weak, resonant coupling between the three modes~\cite{hofer:2017njp,Gonzalez2017,seah:2018}, so that
\begin{equation}
\label{localDissipators}
\LL_j = \kappa_j (1 + n_B^j) \DD[\a_j] + \kappa_j n_B^j\DD[\adag_j].
\end{equation}
Here, $\DD[\hat{L}]\hat{\rho} = \hat{L}\hat{\rho}\hat{L}^\dagger - \tfrac{1}{2}\{\hat{L}^\dagger\hat{L},\hat{\rho}\}$ denotes a Lindblad dissipator, $\kappa_j$ are the cavity decay constants and $n_B^j$ is the Bose-Einstein distribution
\begin{equation}
\label{BEdistribution}
n_B^j = \frac{1}{\ee^{E_j/k_B T_j} - 1}.
\end{equation}
The heat current entering the system from bath $j$ is then defined as
\begin{equation}
\label{heatCurrents}
J_j = \Tr\left \lbrace \H_j \LL_j\hat{\rho}\right \rbrace = \kappa_j E_j \left ( n_B^j - \Tr \left \lbrace \adag_j\a_j\hat{\rho}\right \rbrace \right ).
\end{equation}
In order for the effective Hamiltonian~\eqref{Heff} to remain valid in the presence of dissipation, we require that $\kappa_j \ll E_c,\Delta$, i.e., the cavity linewidths must be much smaller than the sideband frequency and the detuning from electronic resonance. Using Eq.~\eqref{masterEquationCavity}, it is straightforward to verify that Eq.~\eqref{COPperfect}, i.e., $\varepsilon = E_c/E_h$, holds in the steady state when $\partial_t\hat{\rho} = 0$.

\begin{figure}
\includegraphics[scale=0.3]{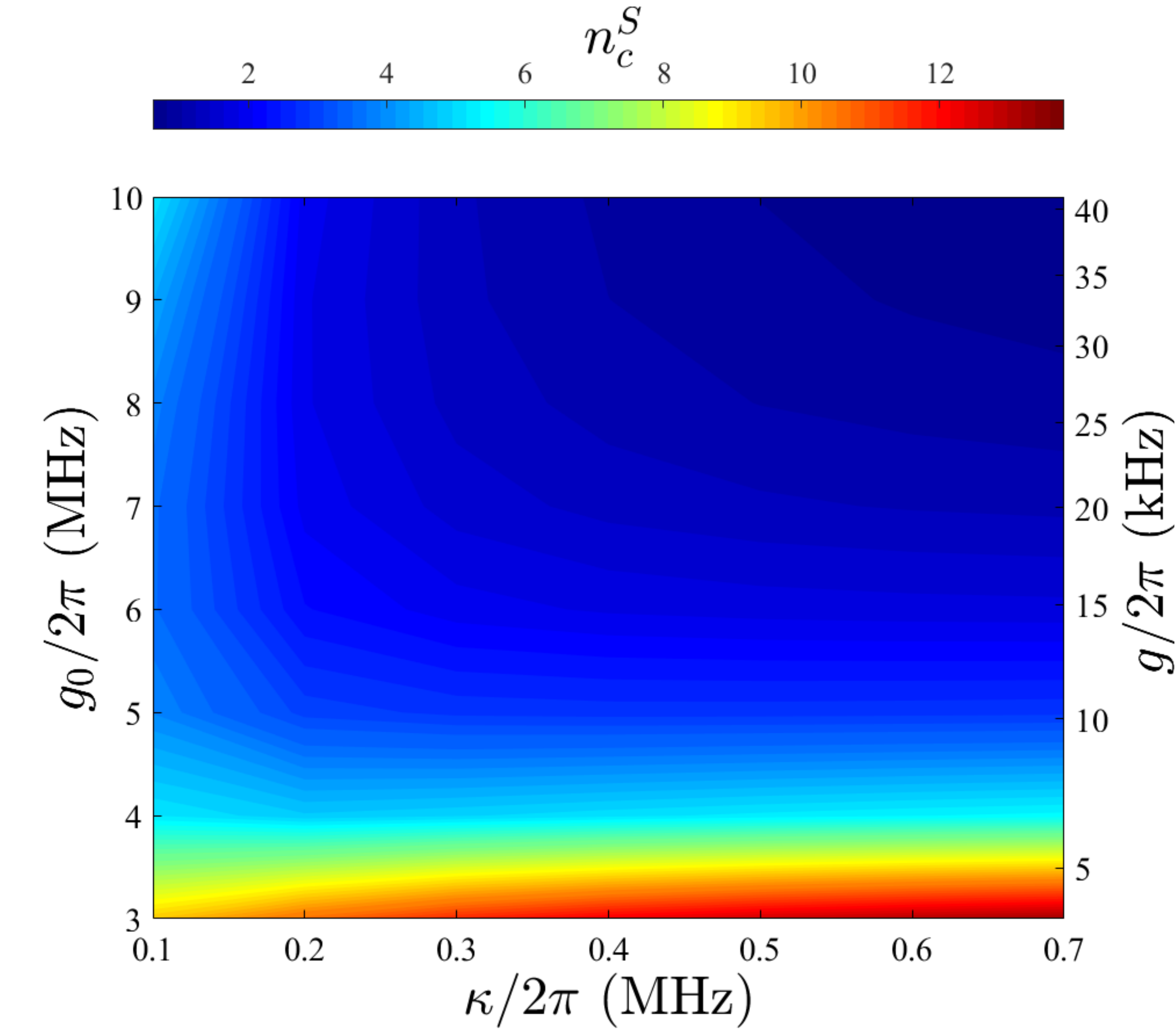}
\caption{Steady-state phonon occupation for the atom-cavity absorption refrigerator, as a function of the (identical) cavity coupling constants $g_h = g_r = g_0$ and decay rates $\kappa_h = \kappa_r = \kappa$. The effective three-body interaction strength $g=-g_0^2\eta/\Delta$ is also shown for comparison. Parameters: $\Delta/2\pi = -100~$MHz, $\Gamma/2\pi = 20~$MHz, $\eta = 0.041$, $\delta = 10~$nm, $\kappa_c n_B^c = 10$~s$^{-1}$, with the remaining parameters given in Tab.~\ref{parameter_table}. Reproduced from Ref.~\cite{mitchison:2016} with modified axis scale and labels. \label{cavity_contour}}
\end{figure}

In Fig.~\ref{cavity_contour} we plot the steady-state occupation of the phonon mode $n_c^S = \lim_{t\to \infty} \Tr[\adag_c\a_c \hat{\rho}(t)]$ for fixed temperatures and several different cavity parameters (see the caption of Fig.~\ref{cavity_contour} and Tab.~\ref{parameter_table}). Our calculation accounts for several experimental imperfections that we have not included in Eq.~\eqref{masterEquationCavity} for the sake of brevity. These include a small, identical misalignment $\delta$ of the trap potential minimum from the node of cavity $h$, and an equal displacement $\delta$ from the anti-node of the cavity $r$. In addition, we assume that hot thermal light impinges on cavity $h$ from only one side, with the other side coupling to the ambient field at temperature $T_r$. Furthermore, we account for dissipative effects due to spontaneous emission, which are suppressed by a factor $\Gamma/\Delta$ relative to the other terms in Eq.~\eqref{masterEquationCavity}. Note that these effects represent heat leaks that degrade performance. See Ref.~\cite{mitchison:2016} for full details of the calculation. 

We choose $T_c = T_r = 300~$K, corresponding to room temperature, while $T_h = 5800~$K, the temperature of sunlight. Despite this seemingly high temperature, the mean number of photons in the cavity is very small, $n_B^h \approx 10^{-3}$, due to the high frequency of the optical mode. Nevertheless, the presence of the cavity greatly enhances the effect of each photon by confining it to a small volume around the atom, and we find that cooling the atomic motion close to its ground state ($n^S_c \sim 1$) is possible so long as the cavity coupling constant is sufficiently large. Note that to actually achieve such temperatures with direct sunlight at the Earth's surface would require optical concentration in order to compensate for the attenuation of intensity with distance from the Sun~\cite{Winston1970,Alicki2015}. Alternatively, one could simulate thermal light using phase-randomized laser light~\cite{Martienssen1964,Arecchi1965} for which very high values $n_B^h\gg 1$ can be obtained, leading to an even larger cooling effect. We also mention that the effective coupling constant $g$ can be enhanced by trapping $N$ atoms simultaneously inside the cavity and cooling their collective center-of-mass motion~\cite{mitchison:2016}. This leads to an $N$-fold increase in $g$, where values $N\sim 500$ are experimentally feasible~\cite{Herskind2009}. 

Converting the steady-state phonon occupation $n_c^S$ to a temperature $T_c^S$ via the relation~\eqref{BEdistribution}, we find $T^S_c\lesssim 1~$mK. (Note that this effective temperature serves only to illustrate the mean energy since the mode is not necessarily in a thermal state.) On the other hand, the cooling power and the COP are exceedingly small (see Tab.~\ref{parameter_table}). Both of these facts can be understood from the very large frequency mismatch between vibrational and optical degrees of freedom. The achievable temperature can be estimated from the virtual temperature, which for $T_h\gg T_r$ reduces to
\begin{equation}
\label{virtualTemperatureApprox}
T_v\approx \frac{E_c}{E_r}T_r.
\end{equation}
Since $E_c/E_r\sim 10^{-8}$ for typical vibrational and optical frequencies, this leads to a very small virtual temperature. On the other hand, since the ideal COP~\eqref{COPperfect} is given by $\varepsilon = E_c/(E_r-E_c)$, it is clear that $\varepsilon$ is very small when $E_c\ll E_r$. This is quite intuitive, since a large quantity of heat $E_h = E_r-E_c$ must be absorbed from the hot bath for every quantum of energy $E_c$ extracted from the cold subsystem. 

We finally address a very important question that we have neglected thus far. Can it really be possible to select the cavity frequencies to differ by such a relatively small amount, $E_r-E_h = E_c \approx 10^{-8}E_r$? Remarkably, the answer is yes, because the cavity frequencies can be controlled to very high precision using methods such as the Pound-Drever-Hall technique~\cite{Black2001}, which requires an auxiliary laser. Importantly, this stabilization can be performed without disrupting the operation of the refrigerator, by using other frequency or polarization modes of each cavity. This indicates that cooling of very low-frequency degrees of freedom is facilitated by a stable frequency reference (the coherent laser field) or, equivalently, an accurate clock.

\section{Inelastic Cooper pair tunneling in circuit QED}
\label{sec:cp}

\begin{figure}[b]
\centering
\includegraphics[width=.95\textwidth]{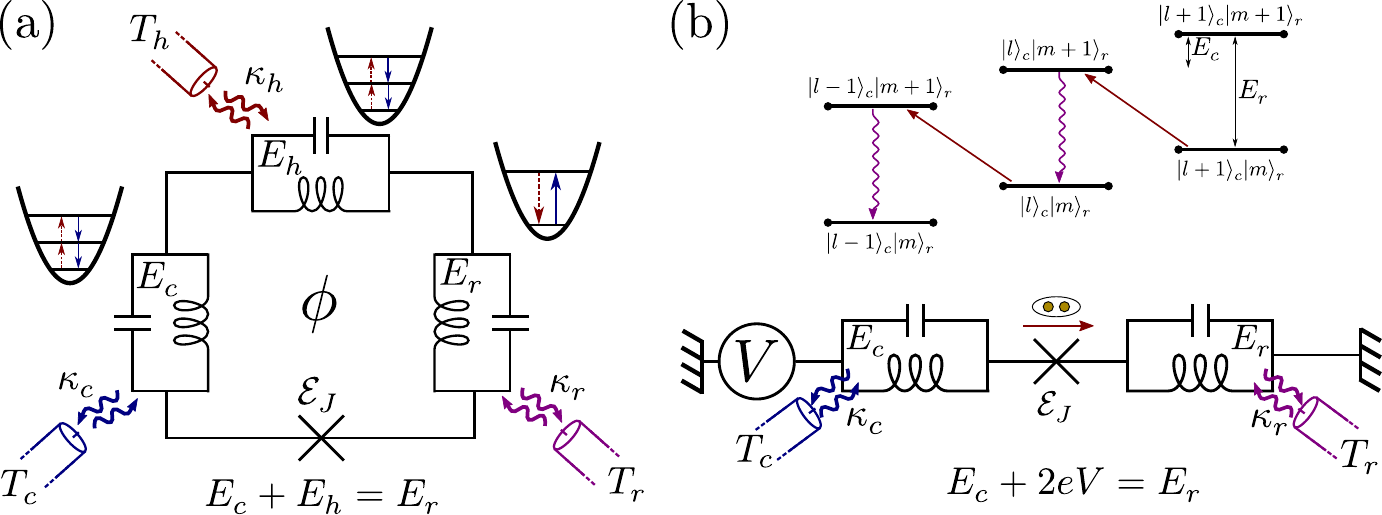}
\caption{${\rm (a)}$ Quantum absorption refrigerator implemented in a superconducting circuit. The subsystems are provided by three $LC$-resonators (with frequency $E_j$) coupled to a thermal bath each (with coupling strength $\kappa_j$). The Josephson junction (with Josephson energy $\mathcal{E}_J$) provides a highly nonlinear interaction between the subsystems. The resonance condition [cf.~\eqr{resonanceCondition}] is key to obtaining an effective three-body interaction [cf.~\eqr{threeBodyInteraction}]. A magnetic flux $\phi$ which pierces the structure can act as an on-off switch. This allows for observing coherence-enhanced cooling in the transient regime. Figure taken from Ref.~\cite{hofer:2016}. ${\rm (b)}$ Implementation using an external voltage source instead of the hot bath as the energy source. The energy quanta of the resonator coupled to the hot bath are replaced by Cooper pairs tunneling across the Josephson junction, gaining or losing the energy $2eV$. The voltage source thereby acts as a coherent external field, providing energy in the form of work. In this implementation, the refrigerator is sideband-cooling a microwave mode. Lower panel taken from Ref.~\cite{hofer:2016prb}.}
  \label{fig:cp_schematics}
\end{figure}

An alternative platform for implementing the quantum absorption refrigerator is provided by superconducting circuits \cite{schoelkopf:2008,vool:2017}, as sketched in Fig.~\ref{fig:cp_schematics}. Within such circuits, $LC$-resonators (where $L$ stands for an inductance and $C$ for a capacitance) with a high quality factor (i.e., with low dissipation) can be designed. These $LC$-circuits can be modeled as standard quantum harmonic oscillators with frequencies in the microwave regime and will provide the subsystems of the refrigerator. A Josephson junction in series with the resonators results in a highly nonlinear coupling between the resonators. The circuit can then be described by the Hamiltonian \cite{armour:2013,gramich:2013,saidi:2001}
\begin{equation}
\label{eq:ham0cp}
\hat{H}(t)=\sum_{j}E_j\hat{a}^\dagger_j\hat{a}_j- \mathcal{E}_J\cos\left(\sum_{j} \hat{\varphi}_j+2eVt+\phi\right),
\end{equation}
where the resonators are labeled by $j$, $E_j$ denote their frequencies, and $\hat{a}_j$ are the corresponding annihilation operators. The Josephson energy of the junction is given by $\mathcal{E}_J$, $V$ denotes an external voltage and $\phi$ is a phase offset that is possibly due to a magnetic field (see below). The flux operators associated with the resonators read
\begin{equation}
\label{eq:fluxes}
\hat{\varphi}_j=\lambda_j(\hat{a}_j^\dagger+\hat{a}_j),
\end{equation}
with $\lambda_j$ being a coupling constant that grows with the impedance of the resonator. 

The Hamiltonian in Eq.~\eqref{eq:ham0cp} is just the sum of the noninteracting resonators, which define the Hamiltonians of the subsystems as $\hat{H}_j=E_j\hat{a}_j^\dagger\hat{a}_j$, plus the Josephson Hamiltonian which is determined by the cosine of the phase difference across the junction \cite{ingold:1992,vool:2017}. Here the phase difference includes the fluxes of the resonators as well as the external voltage. Physically, the Cooper pairs that tunnel across the junction can exchange photons with the resonators to make up for the difference in chemical potential across the junction \cite{averin:1990,ingold:1992,holst:1994,basset:2010,hofheinz:2011}. This results in the highly nonlinear interaction between the resonators.

Writing the cosine as a power series, it is clear that the Hamiltonian includes terms of all orders of $\hat{a}_j$. The versatility of the system stems from the fact that the external voltage can be used to tune different terms in and out of resonance. This versatility has not gone unnoticed, resulting in a number of theoretical proposals based on Eq.~\eqref{eq:ham0cp}. In addition to the thermal machines that are discussed below, these proposals include the stabilization of a Fock state \cite{souquet:2016}, the creation of a single-photon source \cite{dambach:2015}, the creation of nonclassical photon pairs \cite{leppakangas:2013,armour2015josephson,trif:2015}, the generation of bi- and multipartite entanglement \cite{dambach:2017}, and the detection of Majorana fermions \cite{dmytruk:2016}.

On the experimental side, tremendous progress has been made in recent years. The latest achievements include the observation of nonclassical radiation \cite{westig:2017}, paving the way to applications based on entangled microwave photons, and the implementation of a maser \cite{chen:2014} as well as a parametric amplifier \cite{jebari:2018}, demonstrating a high level of control in these systems.

In order to implement the quantum absorption refrigerator, the resonators have to be coupled to thermal baths. This can in principle be done via transmission lines, allowing the heat baths to be far away from each other and thus facilitating the task of maintaining a temperature gradient. As in the previous section, we model the coupling to the heat baths using a standard local Lindblad approach resulting in the master equation
\begin{equation}
\label{eq:lindblad}
\partial_t\hat{\rho} = -i[\hat{H},\hat{\rho}]+\sum_j\LL_j\hat{\rho},
\end{equation}
where the dissipative terms are given in Eq.~\eqref{localDissipators}.

\subsection{Absorption refrigerator}
The quantum absorption refrigerator is implemented with three $LC$-resonators, i.e., $j=c,r,h$ in \eqr{eq:ham0cp}. Furthermore, since no work is involved, the external voltage is set to $V=0$, resulting in a time-independent Hamiltonian. It can be shown that this Hamiltonian is well approximated by 
\begin{equation}
\label{eq:hamfridge}
\hat{H}=\sum_{j=c,r,h}\hat{H}_j+\sin(\phi)\hat{H}_{\rm on}+\cos(\phi)\hat{H}_{\rm off},
\end{equation}
where $\hat{H}_{\rm on}$ describes a three-body interaction of the form of Eq.~\eqref{threeBodyInteraction}, with
\begin{equation}
\label{eq:cpladder}
\hat{L}_j=\hat{A}_j\hat{a}_j,\hspace{1.5cm}g=-8\lambda_c\lambda_r\lambda_h\mathcal{E}_J.
\end{equation}
Here we introduced the Hermitian operators
\begin{equation}
\label{eq:aops}
\hat{A}_j=\ee^{-2\lambda_j^2}\sum\limits_{n=0}^{\infty}\frac{L_{n}^{(1)}(4\lambda_j^2)}{n+1}|n\rangle_j\langle n |,
\end{equation}
with $L_n^{(1)}(x)$ denoting the generalized Laguerre polynomial of degree $n$. Since $[\hat{A}_j,\hat{H}_j]=0$, Eq.~\eqref{loweringOperator} can easily be verified. The approximative Hamiltonian in Eq.~\eqref{eq:hamfridge} is derived under a rotating-wave approximation, where all terms that do not commute with $\sum_j\hat{H}_j$, as well as terms that are higher order in $\lambda_j$, are dropped. For a detailed derivation, we refer the reader to Ref.~\cite{hofer:2016}, where Eq.~\eqref{eq:hamfridge} was compared to the full Josephson Hamiltonian and good agreement was found.

\begin{figure}
\centering
\includegraphics[width=\columnwidth]{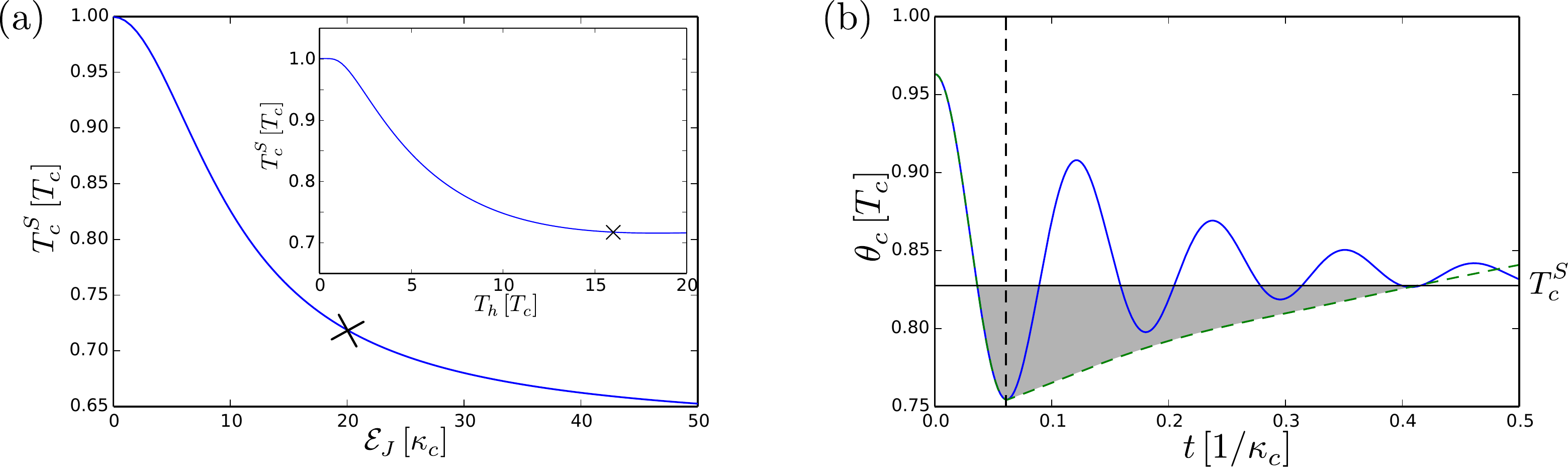}
\caption{Performance of the refrigerator implemented in a superconducting circuit. ${\rm (a)}$ Steady-state cooling. The steady-state temperature in resonator $c$, $T_c^S$, is shown as a function of the Josephson energy. The stronger the interaction between the resonators, the better the cooling. The inset shows the steady-state temperature as a function of the hot bath's temperature. The hotter the hot bath, the stronger the energy source which drives the refrigerator. For large temperatures $T_h$, cooling is reduced by the nonlinear terms in Eq.~\eqref{eq:cpladder}. The cross denotes the point of operation discussed in Tab.~\ref{tab:params}. Parameters: $\kappa_c=\kappa_h=2\pi\cdot0.01\,$GHz, $\kappa_r=2\pi\cdot0.025\,$GHz, $\lambda_c=\lambda_r=\lambda_h=0.3$, $\mathcal{E}_J=2\pi\cdot0.2\,$GHz (inset). Other parameters are given in Tab.~\ref{tab:params}. ${\rm (b)}$ Coherence-enhanced cooling. The temperature in resonator $c$, $\theta_c$, is shown as a function of time. At $t=0$, the refrigerator is switched on. The temperature then oscillates before it reaches $T_c^S$ (blue, solid line). Switching the refrigerator off at the time the temperature is at its first minimum (dashed vertical line) allows for $\theta_c$ to remain below $T_c^S$ for a substantial amount of time (green, dashed line) as illustrated by the grey shading. At $t=0$, $\theta_c$ is below the bath temperature because the system also works as a (poor) refrigerator in the off state (i.e., when $\phi=0$). Parameters are as in ${\rm (a)}$ except for $\kappa_c=\kappa_r=\kappa_h=2\pi\cdot 0.001\,$GHz and $T_h=384\,$mK. Figure taken from Ref.~\cite{hofer:2016}.}
  \label{fig:cp_perf}
\end{figure}

A particularly useful property of this implementation is the fact that the Hamiltonian can be modified by an external magnetic field. This provides the refrigerator with an on-off switch, allowing for the investigation of the transient regime, where coherence-enhanced cooling can be observed. Before exploring the transient regime, we focus on refrigeration in the steady state for $\phi=\pi/2$, i.e., while the fridge is switched on. Figure \ref{fig:cp_perf}~${\rm (a)}$ shows the steady-state temperature of the resonator coupled to the cold bath. As expected, we find a better cooling, the stronger the interaction between the resonators is. As shown in the inset, a higher temperature of the hot bath also increases the cooling. This is due to the fact that the hot bath is the resource of the refrigerator. The hotter it is, the stronger the resource we have at our disposal. Additional parameters which characterize the performance of the refrigerator are given in Tab.~\ref{tab:params} and will be compared to the atom-cavity implementation below. Note that resonator $c$ is strictly speaking not in (although close to) a thermal state. The temperature given here is the temperature of the thermal state with the mean energy of resonator $c$. As in the previous section, the figure of merit is thus the mean energy of resonator $c$. See Ref.~\cite{hofer:2016} for a more detailed discussion on this.

Thanks to the magnetic flux dependence of the Hamiltonian in Eq.~\eqref{eq:hamfridge}, the refrigerator can be turned on and off. For $\phi=\pi/2$, the refrigerator is on, for $\phi=0$, it is off. The Hamiltonian $\hat{H}_{\rm off}$ also describes a refrigerator, but with a coupling strength that goes as $(\lambda_c\lambda_r\lambda_h)^2$, substantially suppressing any cooling. Due to the coherent nature of the three-body interaction, the temperature in resonator $c$ will exhibit a damped oscillatory behavior upon switching on the refrigerator. These oscillations in temperature go well below the steady-state temperature. The on-off switch can be used to cool the resonator below the steady-state temperature by employing the following protocol: The refrigerator is switched on at $t=0$. When the temperature in resonator $c$ reaches its first minimum, the refrigerator is switched off again.
Resonator $c$ will then thermalize with the cold bath on a time scale given by $1/\kappa_c$. For a sufficiently small coupling between the resonator and the cold bath, resonator $c$ will remain at a temperature below the steady-state temperature of the refrigerator for a substantial amount of time. This is illustrated in Fig.~\ref{fig:cp_perf}~${\rm (b)}$. We note that this transient cooling effect crucially relies on the coherent nature of the interaction. While for certain (in particular, finite-dimensional) systems, such a coherent interaction can be seen as a genuine quantum effect \cite{mitchison:2015,brask:2015}, it was recently shown that in the case of harmonic oscillators, a classical theory can capture the coherence-enhanced cooling effect \cite{nimmrichter:2017}. We note that coherence-enhanced cooling in the transient regime was recently observed experimentally in a trapped-ion system \cite{maslennikov:2017}.

\subsection{Sideband cooling of microwaves}

\begin{figure}
\centering
\includegraphics[width=\columnwidth]{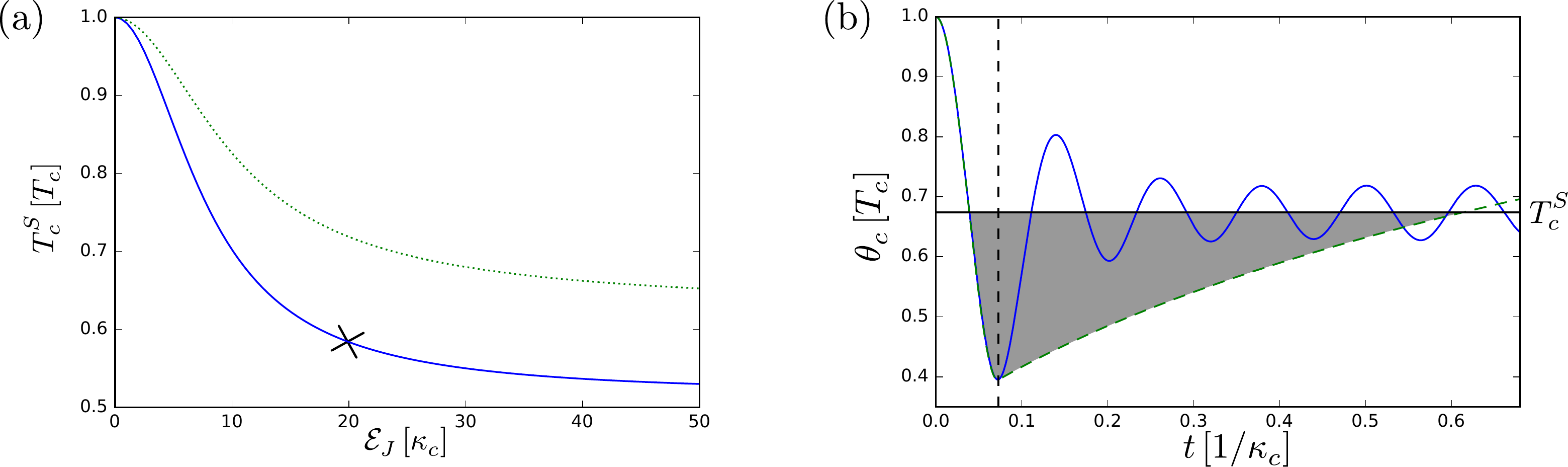}
\caption{Sideband cooling of microwaves. ${\rm (a)}$ Steady-state cooling. The steady-state temperature in resonator $c$, $T_c^S$, is shown as a function of the Josephson energy. For comparison, the result for the absorption refrigerator [cf.~Fig.~\ref{fig:cp_perf}~${\rm (a)}$] is shown as a green (dotted) line. The cross denotes the point of operation discussed in Tab.~\ref{tab:params}. ${\rm (b)}$ Coherence-enhanced cooling. The temperature in resonator $c$, $\theta_c$, is shown as a function of time. Switching the refrigerator off at the time the temperature is at its first minimum (dashed vertical line) allows for $\theta_c$ to remain below $T_c^S$ for a substantial amount of time. Parameters are the same as in Fig.~\ref{fig:cp_perf}.}
  \label{fig:cp_sideband_perf}
\end{figure}

As sketched in Fig.~\ref{fig:cp_schematics}~${\rm (b)}$, the hot bath can be replaced by an external voltage as the energy source for refrigeration. In this case, only two resonators are needed and the Hamiltonian is given by Eq.~\eqref{eq:ham0cp}, with $j=c,r$ and a finite voltage $V$. This Hamiltonian describes Cooper pairs tunneling across a voltage-biased Josephson junction. Due to the voltage, a Cooper pair needs to change its energy by $2eV$ in order to tunnel, which can be achieved by exchanging photons with the resonators. Fixing the voltage to
\begin{equation}
\label{eq:resonancevoltage}
2eV=E_r-E_c,
\end{equation}
a Cooper pair can tunnel by converting photons from resonator $c$ into photons in resonator $r$. In this way an electrical current, powered by the voltage, drives a heat current from the cold bath to the room-temperature bath. Using the resonance condition in Eq.~\eqref{eq:resonancevoltage} and making a rotating-wave approximation results in the approximative Hamiltonian
\begin{equation}
\label{eq:hamsideband}
\hat{H}(t)=\sum_{j=c,r,h}\hat{H}_j+2\lambda_c\lambda_r\mathcal{E}_J\left(\hat{L}_c^\dagger\hat{L}_r\ee^{i2eVt}+\ee^{-i2eVt}\hat{L}_r^\dagger\hat{L}_c\right),
\end{equation}
where $\hat{L}_j$ are given in Eq.~\eqref{eq:cpladder}. For a detailed discussion on the derivation of the last Hamiltonian, we refer the reader to Ref.~\cite{hofer:2016prb}. As before, the coupling to the baths is described by the local Lindblad master equation given in Eq.~\eqref{eq:lindblad}. We note that although the last Hamiltonian includes a time-dependent field, no external time-dependent control is required. The ac Josephson effect converts a time-independent voltage bias into a field oscillating at frequency $2eV$. The performance of the power-driven refrigerator is illustrated in Fig.~\ref{fig:cp_sideband_perf}. While sideband cooling results in a qualitatively similar performance to the absorption refrigerator, a lower steady-state temperature and a higher cooling power can be obtained by using power as the resource (using the same experimental parameters, see also Tab.~\ref{tab:params}). Note however that the COP is the same for the two implementations and given by $\varepsilon=E_c/(E_r-E_c)$ [cf.~Eq.~\eqref{COPperfect}], since both implementations rely on the lossless frequency conversion of photons from $E_c$ to $E_r$.

As for the absorption refrigerator, the cooling process can be switched on and off. In the present case, the refrigerator can be switched off by setting the voltage equal to zero. In this case, tunneling Cooper pairs can not induce photon exchanges and the system is described by the Hamiltonian
\begin{equation}
\label{eq:hamsidebandoff}
\hat{H}=\sum_{j=c,r,h}\hat{H}_j+\mathcal{E}_J\ee^{-2(\lambda_c^2+\lambda_r^2)}\sum_{n,m=0}^{\infty}L_{n}(4\lambda_c^2)L_{m}(4\lambda_r^2)|n\rangle_c\langle n|\otimes|m\rangle_r\langle m|,
\end{equation}
with the Laguerre polynomials $L_n(x)=L_n^{(0)}(x)$. For the parameters used throughout this section, the second term in the last Hamiltonian has hardly any influence on the resonators. As for the previous implementation, coherence-enhanced cooling can thus be observed in the transient regime. Just like in the steady state, using power as a resource results in better cooling for the same system parameters.

\subsection{Quantum thermal machines based on inelastic Cooper pair tunneling}
As discussed above, a number of different effective Hamiltonians can be realized with a Josephson junction coupled to resonators. The ability to tune the desired terms into resonance using externally controlled parameters, such as a voltage and a magnetic field, make this system a promising candidate for implementing different types of thermal machines. In addition to the absorption refrigerator, a heat engine \cite{hofer:2016prb}, a thermometer \cite{hofer:2017}, as well as a machine that extracts work from quantum states \cite{lorch:2018} were investigated.

A heat engine is provided by operating the refrigerator based on sideband cooling in reverse. A temperature bias is then used to drive an electrical current against the voltage bias. Particularly appealing features of this machine are: 1.~Work is unambiguously useful and measurable since it comes in the form of an electrical current. 2.~Because Cooper pairs do not carry any heat, work and heat are carried by different particles (Cooper pairs and photons respectively). These two features make this heat engine a promising candidate to study the intricate interplay of heat and work, as well as their fluctuations, in quantum systems.

The same system can also act as a thermometer. By design, the efficiency of these small thermal machines is uniquely determined by the resonator frequencies. Furthermore, they can reach Carnot efficiency, which is uniquely determined by the bath temperatures. At the Carnot point, the bath temperatures are thus related to the resonator frequencies. Knowing the resonator frequencies as well as all but the smallest bath temperatures allows for a precise estimation of the coldest temperature. This way, an imprecise measurement of a hot temperature can be turned into a precise measurement of a cold temperature. For the system discussed above, the temperature of a microwave resonator can be determined in this way with a precision of $\sim2\,$mK down to temperatures of $\sim15$\,mK.

A single resonator coupled to a Josephson junction can be used to investigate work extraction from quantum states. In this scheme, the state in the resonator is the quantum state of interest and the Cooper pairs are used to extract work from it in the form of an electrical current. Notably, this machine can extract the theoretical maximum of work from all Gaussian and Fock states. The phase difference across the Josephson junction thereby acts as the necessary phase reference for extracting energy stored in the coherences between energy eigenstates. The same system can also be used to investigate the conversion of work stored in a laser field to electrical work. Interestingly, there is a similar efficiency vs.~power trade-off as for heat engines. This implies that work can only change its medium with unit efficiency if it does so infinitely slowly.

\section{Comparison \& outlook}
\label{sec:comparison}
\begin{table}[b]
\def\arraystretch{1.3}
\begin{tabular}{|c|c|c|c|c|c|c|c|c|}
\hline
 Implementation & \,$E_h/2\pi$\, & \,$E_c/2\pi$\, & \,$g/2\pi$\,  &\,  $T_h$ \,&\, $T_{c/r}$\,&\, $T_c^S$\, &\, $J_c$\,&\, $\varepsilon$\, \\\hline\hline
  Atom-Cavity &   \,$810$\,THz\, &\, $5$\,MHz\,&\, $20$\,kHz\,  & \, $5800$\,K\, & \,$300\,$K\, & $1\,$mK\, &$< 0.1\,$yW&\, $< 10^{-8}$\,\\\hline
    Superconductor &   \,$4.5$\,GHz\, &\, $1$\,GHz\,&\, $43.2$\,MHz\, & \, $768$\,mK\, & \,$48\,$mK\, & $36\,(28)\,$mK\, &$0.01\,(0.015)\,$fW\,&\,$22.2\,\,\%$\,\\
\hline
\end{tabular}
\caption{Realistic parameters for the different implementations of the quantum absorption refrigerator. For the superconducting implementation, numbers in brackets are for sideband cooling (i.e., using a power source instead of a hot bath as the resource for cooling).\label{parameter_table} }
\label{tab:params}
\end{table}

In the previous two sections, we discussed two different physical implementations of quantum absorption refrigerators in detail. We have seen that two very distinct physical architectures are well captured by a single framework as they essentially realize the same physics. In particular, the frequency-matching condition in Eq.~\eqref{resonanceCondition} is crucial for obtaining the correct three-body interaction in both architectures. Furthermore, both architectures feature an on-off switch for the interaction. In the superconducting (circuit QED) architecture, this switch is provided by a magnetic field. In the atom-cavity architecture (cavity QED), the atom can be moved from a node to an anti-node (and back). However, in the latter case, the local Hamiltonians of the subsystems are also altered by this process, which can lead to heating unless the switching is performed slowly.

While the underlying cooling mechanism is the same in the two architectures, there are also important differences resulting in a very different performance. In Tab.~\ref{tab:params}, we compare the figures of merit as well as the important system parameters of the two implementations. The most notable difference between the architectures manifests itself in the very different frequency scales. In particular, for the atom-cavity implementation, where the subsystems are provided by different types of physical degrees of freedom, the ratio between the frequencies $E_c/E_h\sim10^{-8}$ is extremely small. While this results in a vanishing efficiency (and requires active stabilization of the cavity frequency), it allows for applying large temperature biases. Together with the small coupling between the cold bath and the system, this allows for cooling a quantum system that is coupled to a room temperature bath down to temperatures of the order of milli-Kelvins. This corresponds to reducing temperature by five orders of magnitude. The superconductor implementation relies on subsystems that are identical in nature, resulting in ratios $E_c/E_h\sim0.22$ relatively close to unity. While this results in large efficiencies, it restricts the amount by which the quantum degree of freedom can be cooled down (here, about a factor of two).

We thus conclude that the superconducting architecture is superior if the aim is to efficiently use resources for cooling and/or if we are interested in cooling down the macroscopic bath. Should we aim at cooling down the quantum degree of freedom coupled to the cold bath, the atom-cavity implementation is the superior one.

Since the implementations discussed here were put forward as proposals, there has been an experimental implementation of a quantum absorption refrigerator based on the motional degrees of freedom of trapped ions \cite{maslennikov:2017}. In contrast to the standard absorption refrigerator, there are no thermal baths involved in this experiment. Instead, the three harmonic oscillators that provide the subsystems of the refrigerator are initiated in thermal states at different temperatures. The system then evolves under unitary dynamics governed by a Hamiltonian including a three-body interaction term. Interestingly, even with so few degrees of freedom, the system shows equilibration in the sense that local observables take on steady-state values at most times. We note that the recurrence time is much larger than any other time scale in this system \cite{nimmrichter:2017}. This opens up opportunities for studying equilibration in closed systems \cite{gogolin:2016}. Furthermore, temperature oscillations dipping below the steady-state value were observed in the experiment in Ref.~\cite{maslennikov:2017}, demonstrating the advantage of a coherent interaction over an incoherent one.

While the quantum absorption refrigerator is based on a quantum mechanical model which includes a coherent interaction, a clear advantage over a classical analogue is not always straightforward to establish (see also Ref.~\cite{levy:book}). If the subsystems of the refrigerator are qubits, a reasonable classical analogue is obtained by replacing the coherent interaction with an incoherent one \cite{mitchison:2015}. In this case, coherence-enhanced cooling in the transient regime provides a clear quantum advantage, where the quantum model outperforms the classical analogue. Furthermore, in a refrigerator based on qubits, entanglement between the subsystems has been connected to an enhanced performance of the refrigerator \cite{brunner:2014}. For refrigerators based on harmonic oscillators, such as the ones discussed here, it is more difficult to determine a genuine nonclassical signature. First, no entanglement between the subsystems has been observed so far. Second, since the classical theory of electrodynamics entails coherence, a classical model for the absorption refrigerator that includes coherence can be constructed. This model captures the coherence-enhanced cooling effect in the transient regime \cite{nimmrichter:2017}. Similarly, one might be able to construct an alternative classical model of the qubit refrigerator in terms of precessing magnetic moments (a qubit can be realized by a spin-half magnetic moment, see also Ref.~\cite{seah:book}), whose oscillatory dynamics could potentially capture the coherence-enhanced cooling effect. This illustrates a general difficulty of demonstrating quantum enhancements in thermodynamics: there exists no universally agreed-upon notion of the ``classical limit'' for a quantum thermal machine. This is in stark contrast to other fields displaying quantum advantages such as computation, where the difference between a quantum and a classical computer is physically and mathematically well defined. 

One possibility for determining an unambiguous quantum signature in absorption refrigerators is provided by considering energy quantization. In classical theories, single degrees of freedom usually exhibit particle-like behavior (which can result in quantization of energy) or wave-like behavior (which can result in coherence) but not both at the same time. With the coherence-enhanced cooling mechanism, we already encountered an observable effect which requires coherence. Most reasonable classical analogues of the absorption refrigerator could be ruled out if energy quantization can be shown to play a crucial role as well. On the surface, the working principle of the quantum absorption refrigerator seems to rely heavily on energy quanta \cite{brunner:2012,correa:2014}. Indeed, it is the fact that a single photon from the hot bath is able to extract a single photon (or phonon) from the cold bath which results in the simple, universal COP. However, in order to certify that the energy passes the refrigerator in quanta, one presumably has to go beyond mean values and investigate fluctuations. In addition to possibly certifying nonclassicality, the study of fluctuations in thermal machines will lead to a deeper understanding of the underlying physics and might lead to novel mechanisms for harvesting energy from fluctuations.

Other avenues to pursue include investigating the role of measurements and possibly feedback on the refrigerator (see, for instance, Refs.~\cite{levy:2016,elouard:2017,mohammady:2017,mancino:book,cottet:book}). Even in autonomous thermal machines, a measurement arguably needs to occur at some time (otherwise, how would one know that something beneficial happened?). This line of research could connect autonomous machines with the thermodynamics of information~\cite{Parrondo2015}. We have also seen that an external frequency reference or clock is useful, both in order to achieve very low steady-state temperatures in the atom-cavity system and in order to take advantage of transient temperature oscillations in the superconducting-circuit fridge. An intriguing question thus arises regarding the energetic and entropic cost of maintaining and using such a clock. Ideally, one would like to analyze the clock within a fully autonomous model in order to ensure fair bookkeeping of resources. While the first steps have been taken in this direction~\cite{malabarba:2015,Barato2016,Woods2016,erker:2017}, much work remains to be done in order to fully understand the thermodynamics of time measurements and time-dependent control of quantum systems.

We conclude this chapter by emphasizing that the quantum absorption refrigerator is an extremely versatile tool: It has already helped to deepen our understanding of thermodynamic processes in the quantum regime and promises to continue doing so in the future. Additionally, the refrigerator actually performs a useful task. Experimental implementations therefore promise to be beneficial on a fundamental as well as a practical level.


\acknowledgements

MTM acknowledges funding from the ERC Synergy grant BioQ and the EU project QUCHIP. PPP  acknowledges  support  from  the  Swiss  National  Science  foundation,  the NCCR  Quantum  Science  and  Technology  (QSIT),  the  Swedish  Research  Council,  and  from  the European  Union’s  Horizon  2020  research  and  innovation  programme  under  the  Marie  Sk\l{}odowska-Curie  Grant  Agreement  No.  796700. We also acknowledge the COST MP1209 network “Thermodynamics in the quantum regime”, which fostered the research described in this chapter.

\bibliography{biblio}

\end{document}